\documentclass[12pt,preprint]{aastex}

\usepackage{lineno}

\shorttitle{EBL_GRB}
\shortauthors{Desai et al.}
\usepackage{calrsfs,euscript,mathrsfs,amssymb}
\usepackage{gensymb}
\usepackage[outdir=./]{epstopdf}

\begin{document}

\title{Probing the EBL evolution at high redshift using GRBs detected with the {\it Fermi}-LAT}
\author{
A.~Desai\altaffilmark{1,2},
M.~Ajello\altaffilmark{1,3},
N.~Omodei\altaffilmark{4,5},
D.~H.~Hartmann\altaffilmark{1},
A.~Dom\'inguez\altaffilmark{6},
V.~S.~Paliya\altaffilmark{1},
K.~Helgason\altaffilmark{7}, 
J.~Finke\altaffilmark{8},
M.~Meyer\altaffilmark{9}
}
\altaffiltext{1}{Department of Physics and Astronomy, Clemson University, Kinard Lab of Physics, Clemson, SC 29634-0978, USA}
\altaffiltext{2}{email: abhishd@g.clemson.edu}
\altaffiltext{3}{email: majello@g.clemson.edu}
\altaffiltext{4}{W. W. Hansen Experimental Physics Laboratory, Kavli Institute for Particle Astrophysics and Cosmology, Department of Physics and SLAC National Accelerator Laboratory, Stanford University, Stanford, CA 94305, USA}
\altaffiltext{5}{email: nicola.omodei@stanford.edu}
\altaffiltext{6}{Grupo de Altas Energ\'ias, Universidad Complutense de Madrid, E-28040 Madrid, Spain}
\altaffiltext{7}{Max-Planck-Institut f\"ur Astrophysik, Postfach 1317, D-85741 Garching, Germany}
\altaffiltext{8}{Space Science Division, Naval Research Laboratory, Washington, DC 20375-5352, USA}
\altaffiltext{9}{Stockholm university, SE-106 91 Stockholm, Sweden}

\begin{abstract}

The extragalactic background light (EBL), from {ultraviolet to infrared wavelengths}, 
is {predominantly} due to emission from stars, accreting black holes and reprocessed light
due to Galactic dust.
The EBL can be
studied through the imprint it leaves, via  $\gamma$-$\gamma$
absorption of high-energy {photons}, in the spectra of distant
$\gamma$-ray sources. 
 {The EBL has been probed through the search for the attenuation it produces in the spectra of
BL Lacertae (BL Lac) objects and individual $\gamma$-ray bursts (GRBs).}
 {GRBs have significant advantages over blazars for the study of {the} EBL 
especially at high redshifts}.
Here we analyze a combined sample of
 {twenty-two} GRBs,
detected by the {\it Fermi} Large Area Telescope {between} 65\,MeV and
500\,GeV.
We report a marginal detection (at the $\sim$2.8\,$\sigma$ level) of
the EBL attenuation in the stacked spectra of the source sample.
This measurement represents a first {constraint} of the EBL at an
 effective redshift of $\sim$1.8. 
We combine our results with prior EBL constraints and conclude that {\it Fermi}-LAT is instrumental to constrain the UV component of the EBL. We discuss the implications on existing empirical models of EBL evolution.

\end{abstract}


\section{Introduction}

Light emitted by stars and accreting compact objects,
through the history of the Universe,
is encoded in the intensity of the extragalactic background light (EBL). 
Cosmic dust in the vicinity of these sources absorbs some fraction of their light and re-emits it 
in the infrared part of the electromagnetic spectrum.
The resulting multi-component spectral energy density is a function of redshift, determined by
cosmological parameters, {stellar initial mass function}, the cosmic star formation rate
history and {the dust content in galaxies} {\citep{2001ARA&A..39..249H,2005PhR...409..361K}}.
Therefore an understanding of the EBL evolution allows us to probe these astrophysical ingredients.
In addition to these standard sources of light, the EBL could also comprise
photons from dark matter particle decay and other potential exotic energy
releases \citep{2012ApJ...745..166M,dom13}.
The evolving EBL in the high redshift domain ($z\gtrsim$6) is of
particular importance as it traces the re-ionization epoch \citep{inoue14}.
Contributions from the first generation of stars (Pop III), might have
originated from very massive stars, which cannot be observed directly
with present day observatories or even with the soon to be launched James Webb Telescope.
These topics have been discussed widely in the literature { \citep{ carr86,dwek05b,raue09,gilmore2012,kashlinsky05,kashlinsky12,inoue13,dwek14}}

Recognizing the importance of {the} EBL and its evolution with redshift, many efforts have been made to 
 {measure its photon intensity}.
Indeed, direct measurements of the EBL are difficult because of the
bright foregrounds like Galactic emission and zodiacal light {\citep{hauser98,matsumoto05,matsuoka11,mattila17}},
 {resulting in estimates of the intensity of the EBL} that are up to a factor of 10 larger than lower limits obtained by integrating the light of
 galaxies resolved in deep surveys \citep{madau00,keenan10,driver16}. 
 {Studies of background fluctuations in the EBL suggest lower, although non-zero, levels
 of unresolved EBL intensity \citep{kashlinsky12,zemcov2014}.}

An  indirect approach of probing the EBL and its redshift evolution
is through the $\gamma$-$\gamma$ absorption {it imprints in the spectra of
distant high-energy $\gamma$-ray sources}. The high-energy part of their spectral energy distributions (SEDs) {is 
attenuated} due to annihilation with background photons ($\gamma$-$\gamma$ $\Rightarrow$ $e^+$-$e^-$ pair creation) {as
discussed by \cite{nikishov61} and \cite{Gould67_2, Gould67}.}
Because of the shape of the pair-production cross section,
$\gamma$ rays (of a given energy) will most likely interact
with EBL photons of given wavelengths: e.g. 
 $\gamma$ rays with  E$\gtrsim$50\,GeV (and from z$\gtrsim$1) are attenuated mainly by photons of the optical-UV background ($>$1\,eV). The total optical depth to a source is then found from a proper  cosmological integration over redshift, which requires an understanding of how the
EBL builds up with cosmic time {\citep{dwek13}.}

This extinction process therefore allows the use of $\gamma$ rays of different energies (and originating
from sources at different redshifts) 
to explore the SED of the EBL and its evolution with redshift.
While the Galactic emissions and zodiacal light constitute a problem for 
direct measurements, they make no difference for the $\gamma$-ray technique as the
mean free path of photons in the MeV to TeV regime is much larger ($>$10\,Mpc) than Galactic or
solar scales \citep{adams}.
Observations {over the $0.2<z<1.6$ redshift range} with the {\it Fermi} Large Area Telescope (LAT) have resulted in the detection 
of the EBL attenuation
in a collective sample of 150 BL Lacertae
objects \citep[BL Lacs, see][]{ebl12}.
Ground-based measurements of low-redshift blazars ($z\lesssim$0.6) in the TeV regime have resulted in
optical depth estimates using High Energy Spectroscopic System (H.E.S.S.), Major Atmospheric Gamma
Imaging Cherenkov Telescopes (MAGIC) and Very Energetic Radiation Imaging Telescope Array System
(VERITAS) data \citep[e.g.][]{hess_ebl12,dominguez13a,biteau2015}.
All measurements in the 0$\lesssim$$z$$\lesssim$1.6 range point to a level of the UV-to-NIR EBL 
that is compatible with that inferred from galaxy counts as estimated by recent
models \citep[e.g.][]{franceschini08,finke10,dominguez11,
stecker12,helgason12,stecker16}.

All measurements of the $\gamma$-ray opacity measured above rely on BL Lacs as probes of the EBL.
Because it has been proposed that the observed $\gamma$-ray absorption may 
be affected by line-of-sight interaction with cosmic rays 
(accelerated in jets of BL Lacs) producing secondary $\gamma$-rays, 
there remain some doubts whether $\gamma$-ray measurements of the EBL using blazars are trustworthy \citep{essey10,essey11}. 
Line of sight
interaction of cosmic rays (accelerated in jets of BL Lacs) with the CMB/EBL would
generate a secondary $\gamma$-ray component, which, being much closer to the observer
would suffer less EBL attenuation and
would lead to underestimation of
the true EBL energy density. 
The detection at TeV energies of  BL Lacs with
unusually hard de-absorbed spectra \citep[e.g.][]{horns12,furniss13} has been interpreted also in this framework. 
 {These possibilities} were discounted by \cite{biteau2015} who find that
the spectra reconstructed after de-absorption are not too hard with respect 
to expectations. 
A similar conclusion was reached by \cite{dominguez2015} who do not find any deviation of the 
predicted EBL attenuation in the LAT blazar data.
In addition to these theoretical uncertainties, the sample of BL Lacs suffers from a significant drop in sample size beyond a redshift of $\sim$1.0.

In this work we overcome these limitations using
the $\gamma$-ray bursts (GRBs) detected by the LAT 
during a 7-year period and for which redshift measurements are available \citep{hartmann07}.
{ The short duration of the bursts} ensures that the observed $\gamma$-ray emission
is generated locally at the source, which renders GRBs clean probes of the EBL.
Furthermore, {GRBs are also observed at much larger redshifts }\citep[i.e., $z=$4.3 for GRB~089016C as reported by][]{080916} thus expanding the study of EBL attenuation to larger distances { \citep[see e.g.][]{kashlinsky5}}. 

This paper is organized as follows: $\S$~\ref{sec:analysis} presents the 
data processing and analysis,
$\S$~\ref{sec:eblanalysis} reports the methodology and results of the EBL study, $\S$~\ref{sec:addtests}
considers systematic effects of the methodology, while $\S$~\ref{sec:discussion} discusses the results.

\section{Data Analysis}
\label{sec:analysis}

\begin{figure*}[ht!]
  \begin{center}
  \begin{tabular}{c}
    \includegraphics[scale=0.6]{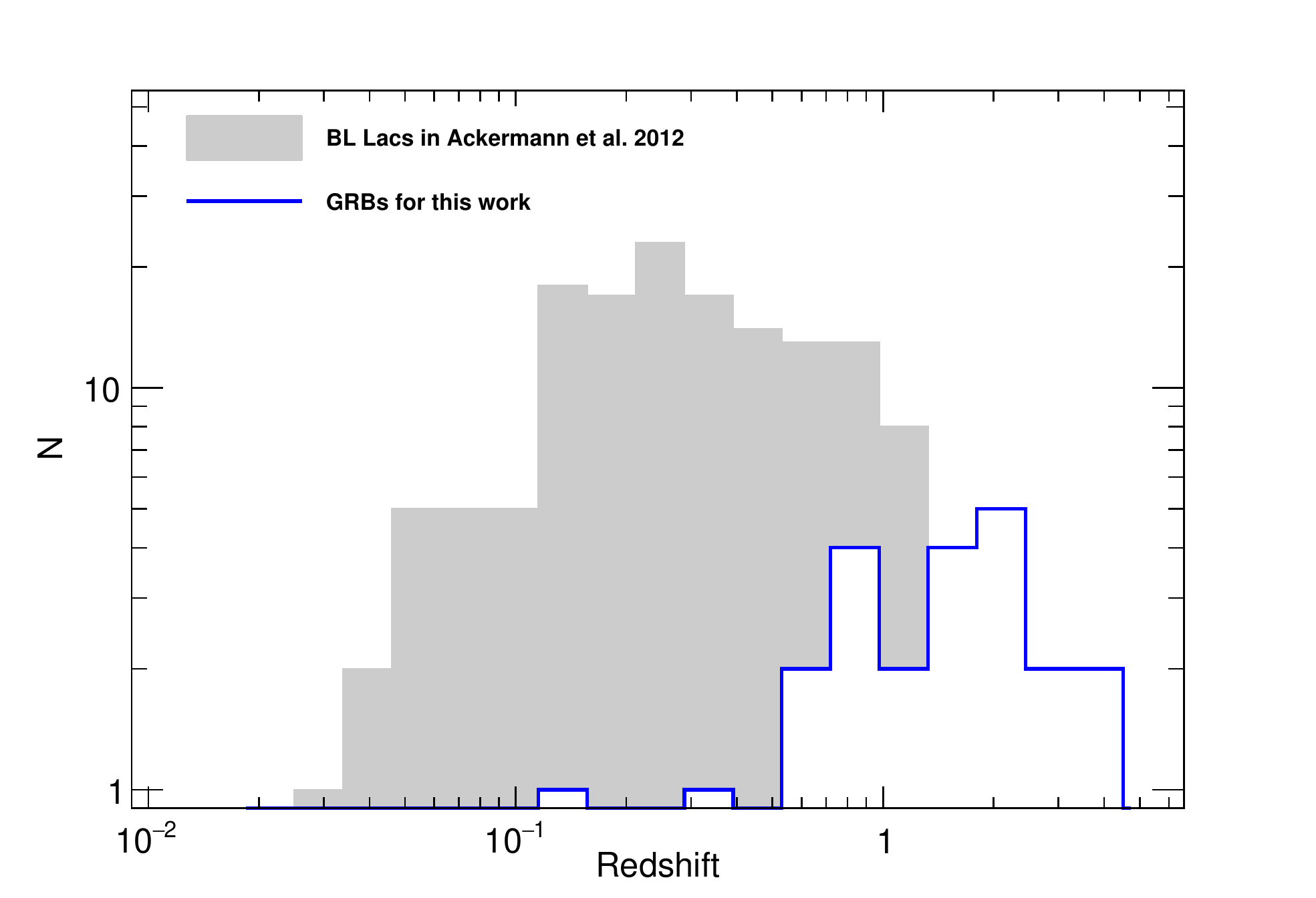} 
  \end{tabular}
  \end{center}
  \caption{Redshift distribution for the sample of {twenty-two} GRBs used here compared to the sample of 150 BL Lacs used in \cite{ebl12}.
\label{fig:redshift}}
\end{figure*}

\begin{figure*}[ht!]
  \begin{center}
  \begin{tabular}{c}
    \includegraphics[scale=0.55]{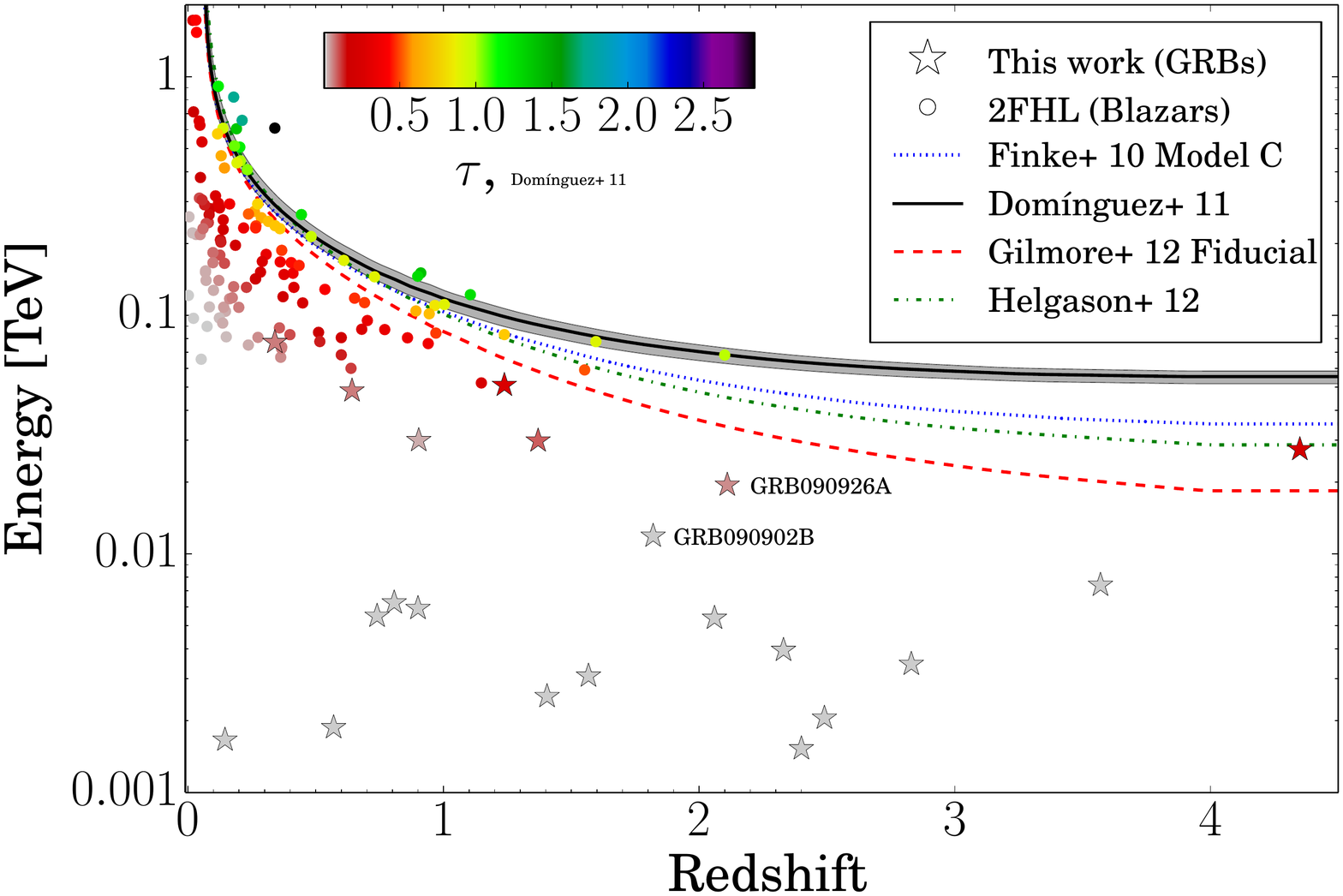} 
  \end{tabular}
  \end{center}
  \caption{ Prediction of the cosmic $\gamma$-ray horizon (i.e. the redshift and energy at which $\tau_{\gamma \gamma}=1$) from different models (see legend) along with the highest energy photons from AGNs and GRBs at different redshifts. The GRBs from our sample are denoted by stars, AGNs by dots while the estimates from EBL models are denoted by lines. The two most constraining GRBs in our analysis are labelled in the plot for reference.
\label{fig:opacity}}
\end{figure*}

There are more than 130 GRBs detected by {\it Fermi}-LAT
\citep{2016AAS...22741601V}, out of which {twenty-two} GRBs measured between September
2008 and June 2015 have an associated redshift measurement, which comprise our source sample.
These GRBs are reported in Table~\ref{tab:list} along with their corresponding parameters. { 
Table~\ref{tab:nphotons} reports the number of photons detected with the {\it Fermi}-LAT at an EBL optical depth greater than 0.1 (obtained using the model of \cite{finke10}-model C and corresponding redshift measurement for each GRB).  In order to show how much the number of photons above a given optical depth varies when the EBL model is changed, we also report the number of photons detected at $\tau>$0.1 using the models of \cite{dominguez11} and \cite{kneiske10} (a more transparent and more opaque model than the one of \citeauthor{finke10} 2010 respectively).}
The redshift distribution for our sample ranges from 0.15 to 4.35 and is shown in
Figure~\ref{fig:redshift} compared to the distribution for BL Lacs from the 
sample used by \cite{ebl12}.
  Figure~\ref{fig:opacity} shows the highest energy photons detected from
these GRBs together with prediction of the cosmic $\gamma$-ray horizon
from different models.


For each GRB, we {extract }transient-class Pass 8 photons detected with the {\it Fermi}-LAT 
between 65\,MeV and 500\,GeV within 10$\degree$ of the source.
The start time (in UTC) and duration of each burst (reported in Table~\ref{tab:list}) is obtained from
the LAT first GRB catalog \citep{lat_grbcat1}, on-line GRB
table\footnotemark{} and individual burst papers {\citep{lat_grb09,palmaetal,kd10,cutoff09,PLspect2013}}.
 {There are no diffuse models available at energies less than 65\,MeV and the effective area of {\it Fermi}-LAT decreases
steeply at low energies, reducing the overall sensitivity. So, to obtain maximum signal strength, we took 65\,MeV as the lower limit for the analysis.}
{The maximal energy }must be $\gtrsim$ 10\,GeV as photons having energy greater than 10\,GeV interact 
with the EBL to produce electron-positron pairs. The universe is transparent
below $\sim$10\,GeV \citep{stecker06} meaning that the measured
spectrum will {be} equal to the intrinsic spectrum for E$\ <10$\,GeV.
To retain sensitivity to EBL attenuation, we adopt 500\,GeV as the
upper limiting energy.

\footnotetext{$http://fermi.gsfc.nasa.gov/ssc/observations/types/grbs/lat\_grbs/table.php$}

The burst data for each GRB are analyzed using {\it Fermi
Science-Tools} (version v10r0p5)\footnotemark{}.
\footnotetext{$http://Fermi.gsfc.nasa.gov/ssc/data/analysis/scitools/$}
These data are
filtered, removing the photons having a zenith angle greater than 105$\degree$,
to limit the
contamination due to Earth's limb {(this analysis is robust against changes in zenith angle cuts\footnotemark{})}.
\footnotetext{Adopting a more stringent zenith angle cut of $85\degree$ produces negligible impact on our analysis.}
The photons collected
by the LAT when it is in the South Atlantic Anomaly (SAA) are also filtered out. 
The spectral analysis of the burst is done by an unbinned likelihood maximization
of a sky model created for each GRB. The sky model consists of a central point source, 
the GRB, whose spectrum is modeled as a power law, and the diffuse (Galactic and isotropic) models.
The Galactic and isotropic models are modeled using the gll$\_$iem$\_$v06.fits and
iso$\_$P8R2$\_$TRANSIENT020$\_$V6$\_$v06.txt templates\footnotemark{}
respectively \citep{lat_diffuse_model16}.
 We {use} the $P8R2$\_$TRANSIENT020$ instrument response function.

\footnotetext{$http://Fermi.gsfc.nasa.gov/ssc/data/access/lat/BackgroundModels.html$}

The Minuit\footnotemark{} optimizer is used to determine the
\footnotetext{$http://lcgapp.cern.ch/project/cls/work-packages/mathlibs/minuit/doc/doc.html$}
best-fit spectral parameters and the error estimate for the unbinned
likelihood maximization analysis. 
GRB spectra are generally described using the ``Band function'' \citep{band93}, which consists of two power laws joined by a exponential cut-off, or a Comptonized model, which
consists of a power law with exponential cut-off \citep{Fermicat}.
According to \cite{Fermicat} and \cite{omodei15}, the ``Band function'' alone is inadequate to 
model GRB spectra over the \,keV-\,GeV energy range observed by {\it Fermi} and a power-law component is required in all
bright LAT bursts to account for the high-energy data ($>100\,$\,MeV).
 {This component may be produced by synchrotron
  radiation resulting in a power-law like  spectrum \cite[as reported
  by \citeauthor{PLspect2013} \citeyear{PLspect2013} and discussed
  also by][]{kd09,kd10,ghisellini10,wang10}}.

We thus approximate the intrinsic spectrum of GRBs with a power law
 {and assess
in $\S$\ref{ssec:expcutoff} how well this assumption works.}

The power law used for our {intrinsic point source spectra} is
given by
\begin{equation}
    {\frac{dN_0}{dE}} = {\frac{N_0(\alpha +1)E^\alpha}{E_{max}^{\alpha+1}-E_{min}^{\alpha+1}}}
\end{equation}

where $N_0$ gives the normalized flux in units of { \,cm$^{-2}$s$^{-1}$MeV$^{-1}$}
between $E_{min}$ and $E_{max}$ taken as 65\,MeV and
500\,GeV respectively, while $\alpha$ is the photon index. 
For the likelihood analysis of each GRB, 3 parameters ($N_0$ and $\alpha$ of
the point source and the normalization of the isotropic diffuse source) are left free to vary
while the rest are fixed.
 {Because of the short time integration of bursts and lack of photons to constrain both background emissions, the Galactic diffuse emission is fixed.}
The log likelihood value obtained from the null case $(LL_{null})$, where the source is not 
present, is compared with the 
log likelihood value obtained from the source model $(LL)$ using the Test Statistic (TS) 
given by $2(LL-LL_{null})$.
The TS value along with the estimated flux  and photon index
are reported, for all GRBs, in Table~\ref{tab:list}.
The source significance, which gives us the confidence level for the detection of each GRB,
is obtained by taking the square root of the TS value
$n_{\sigma}=TS^{1/2}\sigma$ \citep{1996ApJ}.

\section{EBL Study}
\label{sec:eblanalysis}
\subsection{Likelihood Methodology}
\label{ssec:method}

Our EBL analysis aims to find out the attenuation due to the EBL in the spectra of GRBs.
{ To measure the EBL attenuation, in this work we test separately the normalization and shape of optical depth curves predicted by several EBL models. The normalization of the optical depth is tested following a procedure similar to the one of \cite{ebl12} by performing the likelihood ratio test (see also \citeauthor{hess_ebl12} 2013 and \citeauthor{magic16} 2016), while the shape is tested as discussed in Section ~\ref{sec:discussion}. Owing to the limited signal-to-noise ratio of the measurement within the considered energy range, the shapes of most EBL models are found to be similar to each other (also discussed in Section~\ref{sec:discussion}). This similarity makes the LAT data more sensitive to the normalization than to the shape of the models. Moreover this approach is compatible (and allows for an easy comparison) with the method adopted also by e.g. MAGIC, H.E.S.S, and VERITAS \citep{magicebl,hess_ebl12,veritasebl}.}
The EBL
absorption is parametrized as $e^{-b \cdot \tau_{model} }$ where the optical
depth $\tau_{model}$= $\tau(E,z)$ 
is derived by 13 EBL models \cite[see Table~\ref{tab:grbresults} e.g.][]{kneiske04,finke10,dominguez11,
stecker12,helgason12} and depends on the photon energy $E$ and source
redshift $z$ { under} consideration.
This EBL optical depth is scaled to fit the data using the $b$
parameter. The observed spectrum is then given by:
\begin{equation}
    F(E)_{observed} = F(E)_{intrinsic} \cdot e^{-b \cdot \tau_{model} }
\end{equation}
where, {$F(E)_{intrinsic}={{dN_0}/{dE}}$} gives the intrinsic GRB spectrum.

 {A stacking analysis is used to determine the significance of the EBL
attenuation in the observed
GRB spectra and to overcome the limitation of low statistics from single GRB sources.}
In this analysis, the best-fit value of the scaling parameter $b$ is
determined through a simultaneous fit to all GRBs. 
The spectral parameters of each GRB were allowed to vary independently during
the fitting with the exception of $b$ (i.e. the scaled
EBL attenuation is common to all GRBs) {while the parameter of
    the isotropic component is fixed at its best-fitting value (found
    analyzing each single ROI) and
    those of the Galactic model are kept fixed at their nominal, non-optimized, values.}
Therefore, a total of 45 parameters are left free to vary 
(2 parameters for each GRB and 1 parameter given by $b$).

We define two test statistics $TS_0$ and $TS_1$ that are used to
assess, respectively, the significance of the EBL detection and the
inconsistency of a given EBL model with the LAT data. These are
defined as $TS_0=-2 [LL(b_{best fit}) - LL(b=0)] $ and $TS_1=-2
[LL(b_{best fit}) - LL(b=1)]$, where $LL(b_{best fit})$, $LL(b=0)$, and
$LL(b=1)$
are the log-likelihoods of when $b$ was left free to vary, and fixed at
0 and 1 respectively.
The $TS_0$ value is obtained by comparing the null case, which indicates no EBL attenuation,
to the best-fit case. {The significance is calculated using $\sqrt{TS_{0}}\sigma$ which} gives 
the confidence level for the detection of the EBL attenuation.
{ The $TS_1$ value represents a measurement of the significance of the rejection of a given EBL model.} A high value will mean that the model
is rejected as it predicts an attenuation that is larger than observed,
with a significance of the model rejection given by $\sqrt{TS_{1}}\sigma$.
 {We also use the $TS_0$ and $TS_1$ to calculate the p values of a $\chi ^2$ distribution with one degree of
freedom using $p  = \int_{TS}^{\infty} d\chi^2 PDF(\chi^2, DOF = 1) $ where $PDF$ stands for probability density function and $DOF$ stands for degrees of freedom.}

\subsection{Results}
\label{ssec:eblresults}

\begin{figure*}[ht!]
  \begin{center}
  \begin{tabular}{c}
    \includegraphics[scale=0.55]{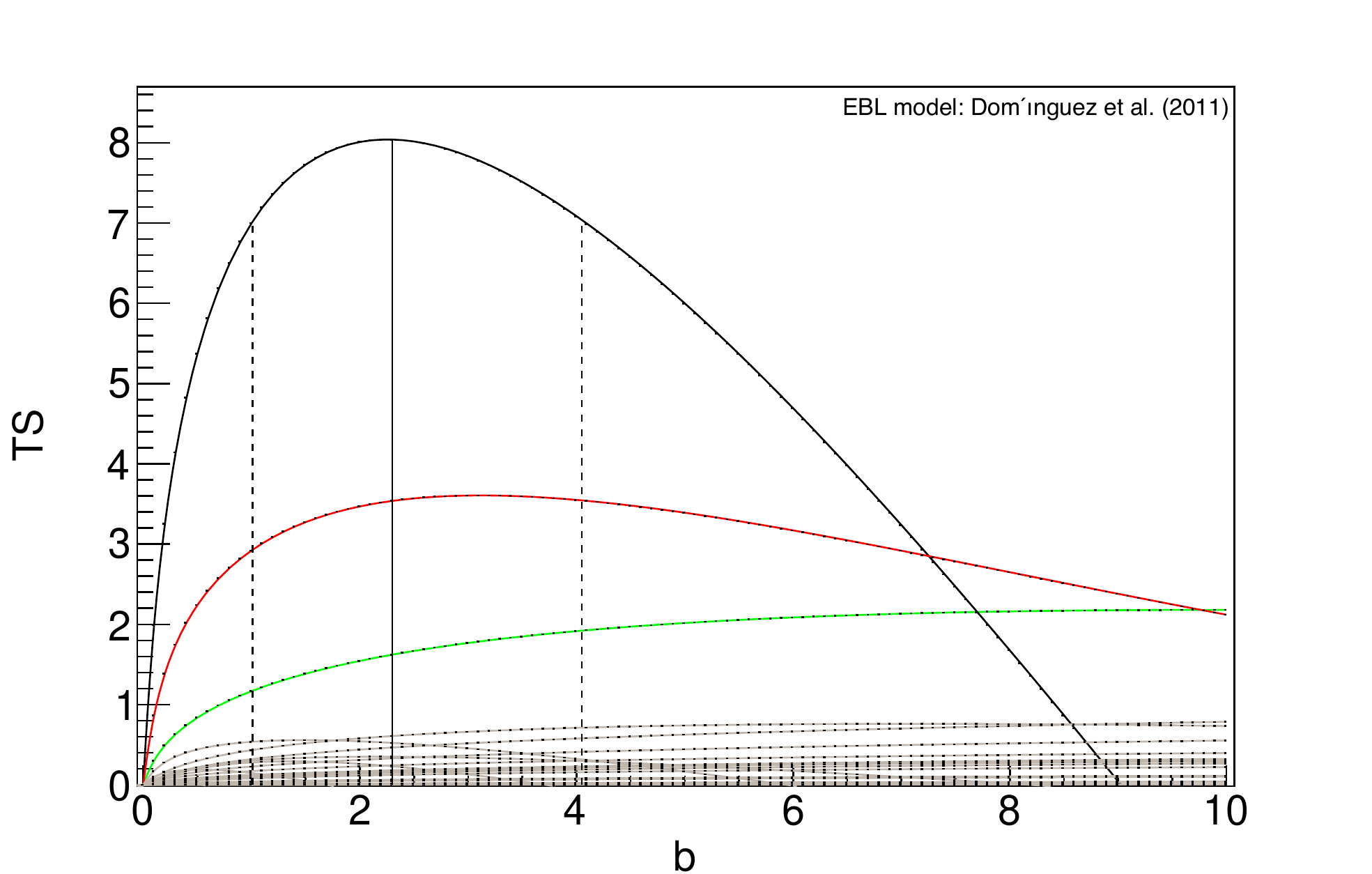}\\ 
    \includegraphics[scale=0.57]{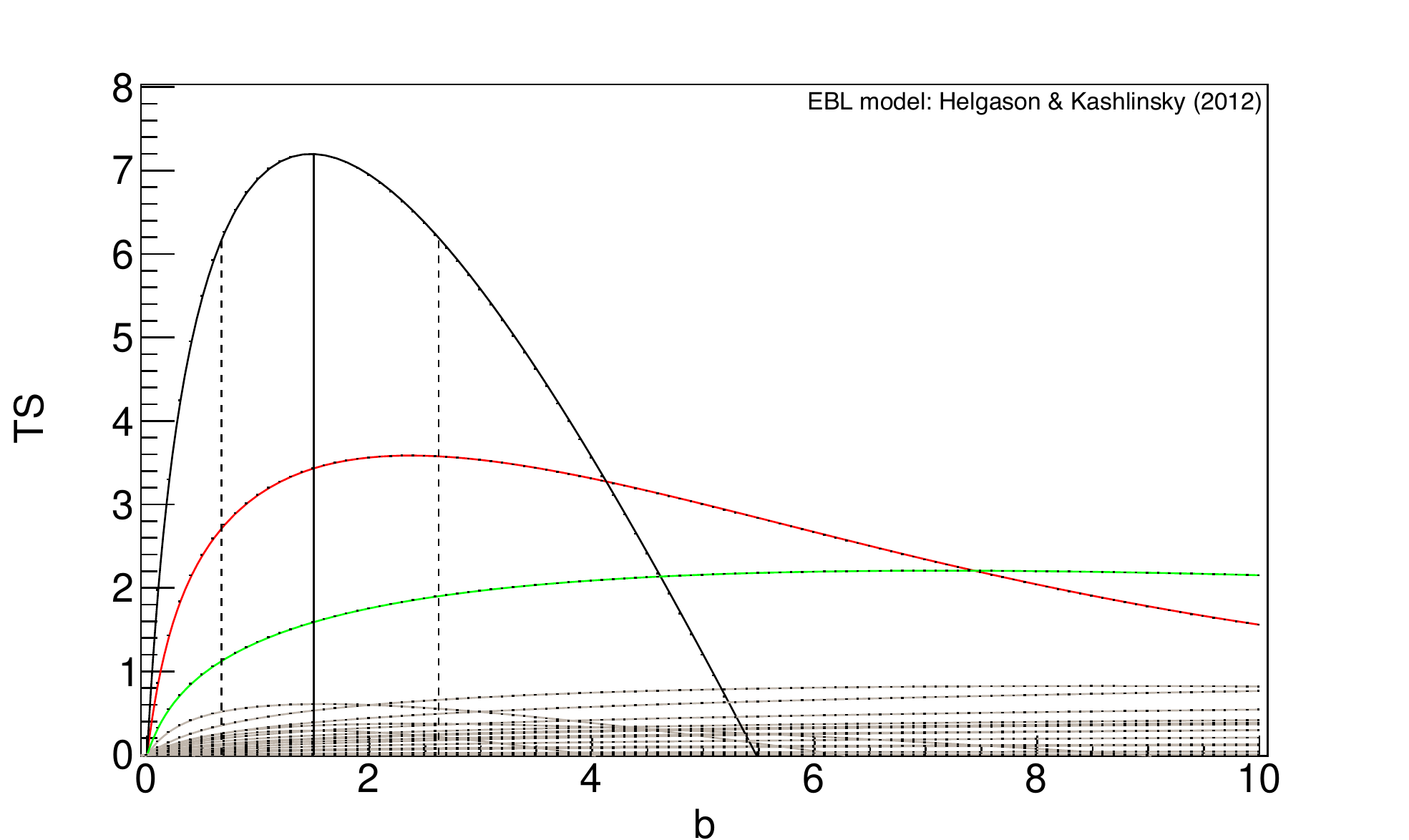} 
\end{tabular}
  \end{center}
  \caption{A combined measurement {(shown by a solid black line)} of the $TS_0$ values of {twenty-two} GRBs as a function of {\it b} is displayed for the { EBL models of \cite{dominguez11} (top) and \cite{helgason12} (bottom)} {along with the measurements for individual GRBs.}  {The solid red and green lines show the maximum contributions to the EBL analysis obtained 
  from GRB090902B and GRB090926A respectively while the solid gray lines show contributions from the remaining 20 GRB sources. The best fit value for the scaling parameter with 1\,$\sigma$ uncertainty values is also shown by the vertical solid and dashed lines respectively.}}
\label{fig:TS}
\end{figure*}

Out of the 13 EBL models tested, the EBL analysis discussed in $\S$~\ref{ssec:method} gave a maximum $TS_0$ value of 8.04 for the EBL model
of \cite{dominguez11} with a best-fit value (with 1\,$\sigma$ uncertainty) of 
$b=2.21_{-1.83}^{+1.48}$.
This rules out the absence of EBL attenuation (${\it b}=0$) at $\sim$2.8\,$\sigma$ {($p= 4.6 \times 10^{-3}$)}.
The plot of $TS_0$ for different {\it b} values obtained using the \cite{dominguez11} model is shown in Figure~\ref{fig:TS}. { Note that the major contribution to the TS comes from GRB090902B and GRB090926A. If these two bursts are excluded from the analysis, we obtain a $b=1.3_{-1.21}^{+1.91}$ and $TS_0=3.04$ for the model of \cite{dominguez11}.}  

The $TS_0$ and $TS_1$ values {along with the $p_0$ and $p_1$ values}, which show the EBL detection and model rejection respectively, for
all the EBL models tested in this analysis are reported in Table~\ref{tab:grbresults}. { We also report the difference between the significance of detection ($TS_0$) and the significance of rejection ($TS_1$). Using the definitions of $TS_0$ and $TS_1$ it is easily seen that their difference will be given by $\Delta TS=-2 [LL(b=0) - LL(b=1)]$. $\Delta$TS represents the improvement in the fit when the nominal (for a given EBL model) estimate of the EBL attenuation is used with respect to the case of no EBL attenuation. A higher value will imply a more significant detection of the EBL at the level nominally derived by the model being tested.}
The EBL models accepted by our analysis are the models having $TS_1$ less than 9, meaning that the model is 
accepted within a 3\,$\sigma$ confidence level. So all the EBL models shown in Table~\ref{tab:grbresults} are compatible with the {\it Fermi}-LAT
GRB data.
For most of the models, the average $TS_0$ is around $\sim$7.3.

\section{Tests for Systematic Effects}
\label{sec:addtests}

\subsection{Intrinsic Spectral Curvature}
\label{ssec:expcutoff}

A spectral break was first {seen in} GRB~090926A 
at a cut-off energy of $\sim$1.4\,GeV \citep{cutoff09}. Recently, \cite{tang2015}
found 6 GRBs showing similar spectral features with cut-off energies
ranging from $\sim$10 to $\sim$500\,MeV ({much lower than
  the energy at which EBL attenuation takes place}).
To assess the impact of intrinsic spectral curvature on our EBL
analysis, we performed {a series of tests} modeling the intrinsic
source spectrum with a power law with an  an exponential cut-off
component, modeled as $e^{-E/{E_C}}$, 
dependent on cut-off energy ($E_C$). The individual source spectrum used for all the GRBs in the 
likelihood fit is given by:
\begin{equation}
    \frac{dN}{dE}=N_0 \left ( {\frac{E}{E_0}} \right )^{\gamma} \exp \left (- {\frac{E}{E_c}} \right )
\end{equation}
where $N_0$ is the normalization in units of $\,$cm$^{-2}$s$^{-1}$MeV$^{-1}$, $\gamma$ is the index, $E_0$ is the scaling energy fixed 
at 200\,MeV and $E_C$ is the cut-off energy. 
In the source spectrum, $N_0$, $\gamma$ and $E_C$ are left free to vary while for the isotropic diffuse
source, the normalization parameter 
 is left free.

{In the first test}, EBL attenuation  is included
at the nominal value using \cite{finke10}, model-C, {owing to the low uncertainty and high $TS_0$ values obtained from our analysis.}
The scaling parameter ($b$) for the EBL 
model is fixed at 1.
So, in all, 4 parameters are optimized for
each GRB. The maximum likelihood is compared with the likelihood obtained by fixing the cut-off energy at 
3\,TeV, which is outside the {\it Fermi}-LAT energy range and is thus equivalent to
having no cut-off in the GRB spectrum (i.e. a simple power-law spectrum).
The Test statistic value obtained from this comparison is denoted by $TS_C$ and is used 
to evaluate the presence of a cut-off in the GRB spectra.

The $TS_C$ value for GRB~090926A is found to be $0.7$ from our analysis, 
which results in a null detection of curvature in the integrated spectrum.
This result is different from \cite{cutoff09} because we used a longer time interval (4889\,seconds) for the GRB sample as compared to the
``prompt'' interval (3-21\,seconds) used by \cite{cutoff09}.
A $TS_C$ value greater than 1.9 is found for only two of the {twenty-two} GRBs in our sample.
GRB~120624 has a $TS_C$ value of 
3.24 and a best-fit value of 1.23 GeV for the cut-off energy and GRB~131108 has a $TS_C$ value of 
4.02 and a best-fit value of 1.13 GeV. The cut-off energies {found for both
GRBs} are significantly
lower than the energy at which EBL attenuation takes place, {and
modeling these two sources as exponentially absorbed power laws has
negligible impact on the significance of the detection of the EBL
attenuation reported in Table~\ref{tab:grbresults}.}

{Secondly, we repeated the above test, adopting an energy range
  that is restricted for every GRB so that the EBL  attenuation is
  negligible\footnote{Each spectrum was fitted up to a maximum energy
    that is derived from each GRB when the attenuation, as predicted
    from \cite{finke10}, is negligible ($<5$\,\%).}. In this way our
  analysis of the curvature of the GRB intrinsic spectra is not
  affected by EBL attenuation. This confirms the results of the
  previous analysis, deriving a $TS_C$ of 3.54 and 4.16 for GRB~120624
  and GRB~131108. Again modeling the spectra of these sources to include
  the exponential cut-off has negligible impact on the EBL.
}

{Thirdly, if the curvature of the intrinsic spectrum were not well
  modeled (by e.g. neglecting exponential cut-offs), this effect would
  be visible as a shift to lower values of the best-fitting
  $b$ parameter as a function of increasing minimum energy adopted in the analysis. 
We thus repeated the entire analysis adopting a minimum energy of
1\,GeV (instead of 65\,MeV) and measured a $TS_0$=5.9 and
$b=1.20^{+1.50}_{-0.85}$ for the \cite{finke10} model, which is in very good
agreement with the results in Table~\ref{tab:grbresults}. This again
shows that modeling the intrinsic GRB spectra with a power law is a
reasonable assumption and that intrinsic curvature, if present, is not biasing the
result of this analysis.}

 {Finally, we also computed the $TS_0$ and $b$ values 
for the \cite{finke10} model 
modeling all the GRB  intrinsic spectra with an exponentially cut-off
power law.
A $TS_0$ of $0.87$ with a $b=0.53$ was obtained, which is significantly lower 
than the result found using a simple power-law model as an
intrinsic spectrum for all GRBs (See Table~\ref{tab:grbresults}). 
However this model employs {twenty-two} additional free parameters (a
cut-off energy per source) while producing a similar log-likelihood as
the EBL absorbed power-law model. {Model simplicity leads us to
  prefer the scenario where the power-law emission of GRBs is attenuated by
  the EBL (a phenomenon already observed in BL Lacs) rather than a
  more complex intrinsic spectrum.} This leads us to
conclude that for the EBL analysis using GRBs, a simple power-law
spectrum is { a reasonable assumption and it is statistically
  preferred, globally, over an exponentially cut-off power-law spectrum.}

\subsection{Time resolved spectral analysis}
\label{ssec:expcutoff}

GRBs are known to display substantial spectral evolution during the prompt phase
\citep[][]{zhang11,1FHL}.
This calls for an additional time-resolved spectral analysis to justify the
usage of time-integrated spectra for the detection of EBL attenuation \citep{guiriec17}. 
We again use \cite{finke10}-model C as the EBL model
for this test.
 {Again} we choose GRB~090902B for this test owing to its relatively high contribution to the
$TS_0$ value. 
The spectrum of GRB~090902B is created for 7 separate time binned
intervals obtained from \cite{090902}. We use a simple power law to model the intrinsic spectrum 
for each time bin.
The process discussed in Section~\ref{sec:analysis} and \ref{ssec:method} is 
followed to obtain $TS_0$ as a function of $b$ for each time bin.
These results are stacked together to obtain a final combined value of $TS_0= 3.4$
corresponding {to} a best fit {$b=1.9^{+3.9}_{-1.4}$ }in agreement to the time-integrated result of $TS_0= 3.5$ and $b=1.8^{+2.8}_{-1.3}$
obtained from Section~\ref{sec:eblanalysis}.
This agreement shows that using time integrated spectra of GRBs does
not have any impact on the detection of the EBL attenuation.

\section{Conclusion}
\label{sec:discussion}

The interaction of $\gamma$ rays from sources at cosmological distances
(e.g. GRBs, blazars, radio galaxies and star forming galaxies)
with EBL photons creates electron-positron pairs causing absorption of $\gamma$ rays
\citep{stecker06}. Using {\it Fermi}-LAT we searched for the imprint of
the EBL in the spectra of {twenty-two} GRBs detected by the LAT and for which redshift
measurements exist. The low number of photons detected from each single GRB
at high energy, predominantly
due to the steep decline (with energy) of the LAT effective area, renders the detection of
the EBL attenuation in the spectrum of a single source challenging.
To overcome this, we analyze the combined set of GRB 
spectra (stacking) which allows us to reject the null hypothesis 
of no EBL attenuation at $\sim2.8\,\sigma$
confidence.

\begin{figure*}[ht!]
  \begin{center}
  \begin{tabular}{c}
    \includegraphics[scale=0.6]{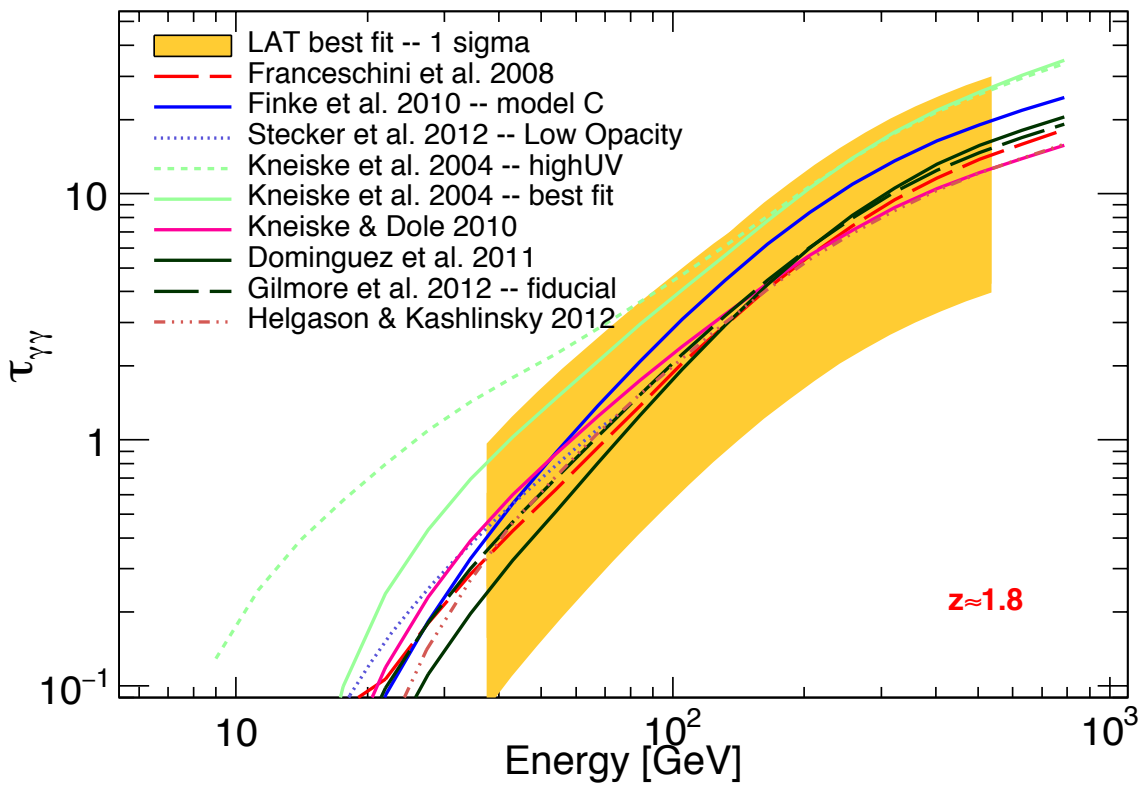} 
\end{tabular}
  \end{center}
  \caption{Constraint on the optical depth at a redshift of $z\approx1.8$, at $1\,\sigma$ confidence
    level $(68\%)$, derived for our GRB sample, compared with model estimates.
 {The models of \cite{franceschini08} and \cite[][high and low opacity]{stecker12}, not included in the numerical analysis (mentioned in Sec.~\ref{sec:eblanalysis}), are included in the figure for completeness.}}
\label{fig:Tau}
\end{figure*}

The constraint on the $\gamma$-ray optical depth as derived from this
analysis is reported in Figure~\ref{fig:Tau}. 
We report this constraint for an effective redshift of $\sim$1.8. 
This value is derived by separating the source sample into two redshift bins and finding the value of the redshift separating the bins for which the $TS_0$ is similar in both bins.}
This helps us to identify the effective redshift based on the 
contribution from each GRB. {  Moreover, dividing the source sample into redshift bins of $0<z<1.8$ and $1.8<z<4.35$, $TS_0= 2.45$ for $0<z<1.8$ and $TS_0= 5.78$ for $1.8<z<4.35$ are obtained, while dividing it into bins of $0<z<1.9$ and $1.9<z<4.35$ gives $TS_0= 5.82$ and $2.18$ respectively. This additional test shows that the effective redshift of our sample is $z\approx1.8$. Also, if GRB090902 at redshift 1.82 is removed from the sample, then the TS values for both the redshift bins are close to equal.}
This effective redshift is slightly higher than the sample average of 1.63 reflecting
the leverage gained by {the high-redshift sources in our sample}.
Figure~\ref{fig:Tau} demonstrates that all the recent EBL models that are in 
agreement with
galaxy counts are also in agreement with the {\it Fermi}-LAT constraint.
The $\gamma$-ray horizon ($\tau$=1) at this effective redshift {occurs} in the range
40 to 180\,GeV, consistent with the range found by \cite{dominguez11} and \cite{2fhl}.
 {As the GRB results are found to be
consistent with those derived for BL Lacs, we conclude that secondary $\gamma$-rays
are not important for moderate optical depths ($\tau\sim1$), as also argued by \cite{biteau2015} and \cite{dominguez2015}.}

The constraints reported in our analysis can also be combined with those reported
by \cite{ebl12} that relied on 150 BL Lacs. These are reported in
Table~\ref{tab:cresults}. While the baseline model of
\cite{stecker06} and the ``high-UV'' model of \cite{kneiske04} were
already found inconsistent with the {\it Fermi}-LAT BL Lac data, we now find
that also the ``best-fit'' model of \cite{kneiske04} is ruled out at {the}
$3\,\sigma$ level when compared to the combined {\it Fermi}-LAT GRB and BL Lac data.  

\begin{figure*}[ht!]

  \begin{center}

  \begin{tabular}{c}
    \includegraphics[scale=0.6]{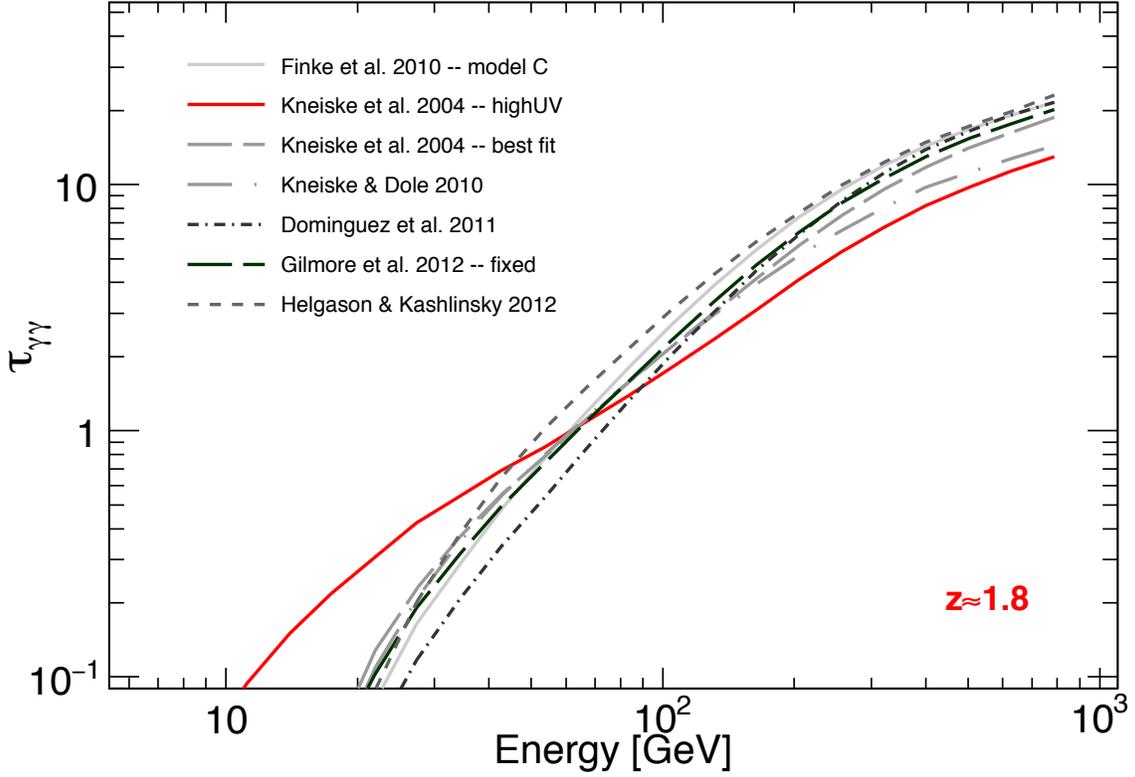}
\end{tabular}

  \end{center}
  \caption{EBL models renormalized to fit the {\it Fermi}-LAT data. The high-UV model of \cite{kneiske04} can be rejected (at 3\,$\sigma$ level) on the basis of the shape of its predicted optical depth curve.}
\label{fig:renormalized}

\end{figure*}

Photons of energy $\lesssim$100\,GeV and from redshift $z>1$ interact
preferentially with
photons of the UV background.
These deviations are appreciated in Figure~\ref{fig:renormalized}, which shows
the models of Table~\ref{tab:cresults} renormalized to fit the {\it Fermi} 
data. It is apparent that all best-fitting (renormalized)
models occupy a narrow region of the $\tau\ vs $ energy plot.
The optical depth curve predicted by the ``high-UV'' model of \cite{kneiske04}
has a shape which is
markedly different than the rest of the models, over-predicting the
optical depth at $<$60\,GeV and under-predicting it above that energy.
This clearly shows that the {\it Fermi}-LAT offers the 
capability to probe the UV background at redshifts $\sim$2, a cosmic epoch
during which the star formation rate density was near maximum
\citep[][]{madau96,bouwens15}.

\begin{figure*}[ht!]
  \begin{center}
  \begin{tabular}{c}
    \includegraphics[scale=0.6]{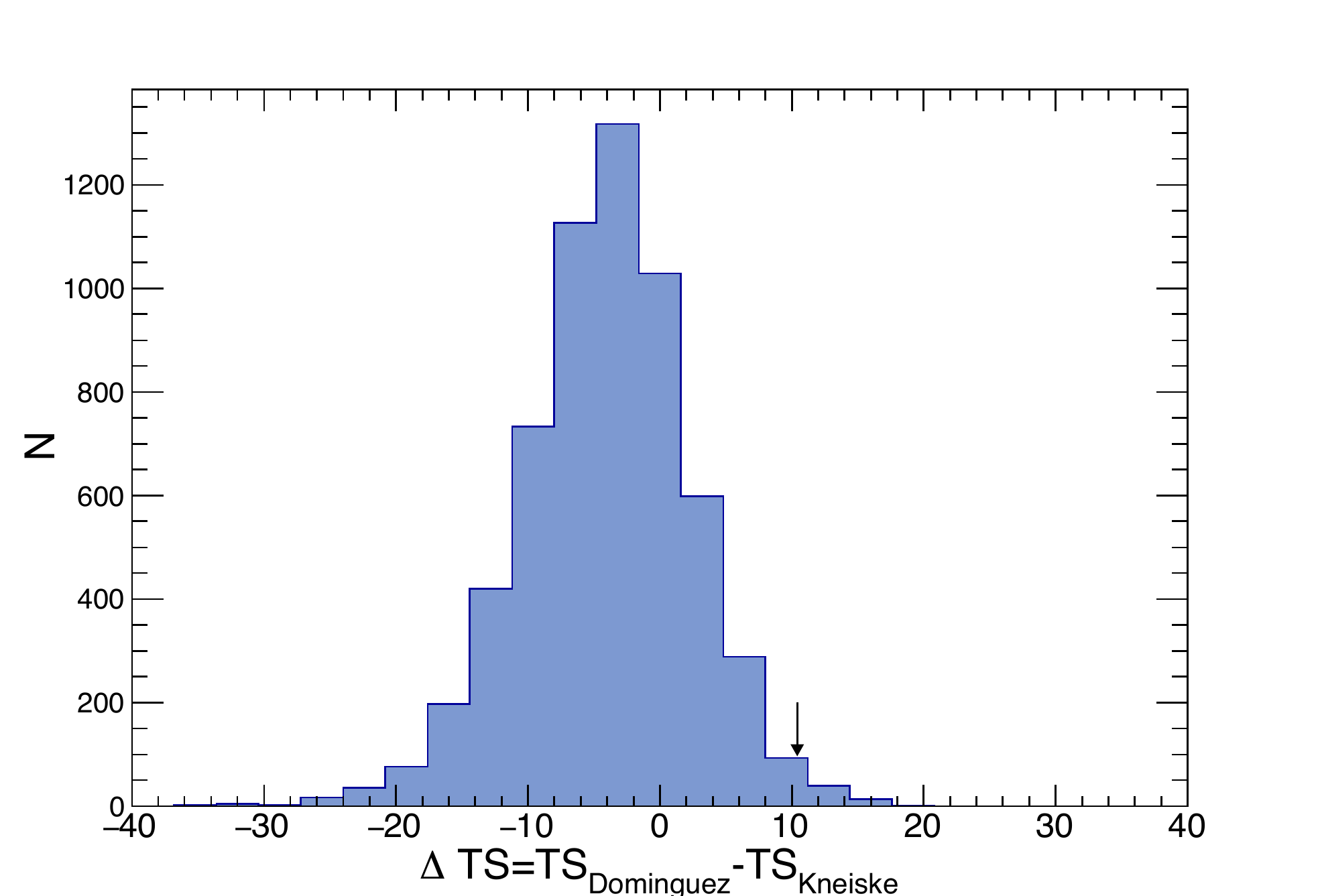} 
\end{tabular}
  \end{center}
  \caption{Distribution obtained from simulations of the $\Delta$TS, when comparing the TS$_0$
    of two different models, in the null hypothesis regime.  In this
    case the simulation adopted the high-UV EBL model of
    \cite{kneiske04}. The arrow
    shows the $\Delta$TS=10.4 value observed in the real data (see Table~\ref{tab:cresults}).} 
\label{fig:nonNested}
\end{figure*}

So far studies have been limited to re-normalizing the EBL models to fit $\gamma$-ray data
\citep[][]{ebl12,hess_ebl12}. This analysis shows that the shape
of the optical depth curve of some models may { be} better than others, even when renormalized 
to fit the LAT data. For example, the \cite{kneiske04}$-–$high UV model implies a
significantly different shape, particularly in the UV (and
correspondingly 10-50\,GeV), as can be
seen from Figure~\ref{fig:renormalized}. In our analysis we allowed every model to be rescaled
by a wavelength-independent factor. Because of the SED shape differences, some models
produce significantly better fits than others even after one allows for different renormalization
factors. This indicates that the analysis presented here is sensitive to the energy dependence of
the EBL thus providing a valuable diagnostic tool. This can be assessed by taking differences
of $TS_0$ values in Table 3. For example the shape (not the normalization) of the optical depth
curve as derived by \cite{dominguez11} is better than the one of the “high-UV” model
of \cite{kneiske04} at {$\Delta$TS$_0$=10.4. However, because the
models are not nested, one needs to calibrate the probability
of observing $\Delta$TS$_0$=10.4, or larger, by chance. We used Monte Carlo simulations of a set of 22 GRBs whose
spectra have been attenuated by the EBL as predicted by the high-UV
model \cite{kneiske04}. Figure~\ref{fig:nonNested} shows the
distribution of $\Delta$TS$_0$ defined as the difference between the
TS$_0$ produced with the \cite{kneiske04} high-UV model and the
\cite{dominguez11} model. We derive that a $\Delta$TS$_0>10.4$ is
observed in $\sim1$\,\% of the cases corresponding to a 3\,$\sigma$
evidence that the shape of the optical depth is better represented by
the \cite{dominguez11} model rather than the high-UV model of
\cite{kneiske04}. This and Figure~\ref{fig:renormalized} show that the LAT is mostly sensitive to the EBL in the UV band, which is
traditionally a very difficult component to model and understand because of
the absorption of light in star-forming galaxies \citep{helgason12}.
}

 {We have shown for the first time that a combined sample of GRBs 
can be used as an excellent probe of the EBL.}
The analysis presented here is based on the relatively small sample of {twenty-two} GRBs with 
known redshifts. However, if we scale the significance of the EBL attenuation by the number
of sources, GRBs appear to have more constraining power than the BL
Lacs used in \cite{ebl12}. This is due to their more simple intrinsic spectrum and high
signal-to-noise spectra that are accumulated over a very short time,
as well as higher
redshift as compared to {BL Lacs in} \cite{ebl12}. 
 {Thus, it is desirable to extend our analysis to a larger burst sample 
underlining the importance of obtaining redshift determinations for 
 {future} GRBs.}

\acknowledgments
The authors acknowledge the comments of the referee. We acknowledge the support of NSF and NASA through grants AST-1715256 and 80NSSC17K0506 respectively.
 The \textit{Fermi} LAT Collaboration acknowledges generous ongoing support
from a number of agencies and institutes that have supported both the
development and the operation of the LAT as well as scientific data analysis.
These include the National Aeronautics and Space Administration and the
Department of Energy in the United States, the Commissariat \`a l'Energie Atomique
and the Centre National de la Recherche Scientifique / Institut National de Physique
Nucl\'eaire et de Physique des Particules in France, the Agenzia Spaziale Italiana
and the Istituto Nazionale di Fisica Nucleare in Italy, the Ministry of Education,
Culture, Sports, Science and Technology (MEXT), High Energy Accelerator Research
Organization (KEK) and Japan Aerospace Exploration Agency (JAXA) in Japan, and
the K.~A.~Wallenberg Foundation, the Swedish Research Council and the
Swedish National Space Board in Sweden.
 
Additional support for science analysis during the operations phase is gratefully
acknowledged from the Istituto Nazionale di Astrofisica in Italy and the Centre
National d'\'Etudes Spatiales in France. This work performed in part under DOE
Contract DE-AC02-76SF00515.

A.D. thanks the support of the Juan de la Cierva program from the Spanish MEC.

{\it Facilities:} \facility{Fermi/LAT}

\bibliographystyle{apj}
\bibliography{biblio.bib}

\setcounter{enumi}{1}

\begin{deluxetable}{clcccccccc}
\tablewidth{0pt}
\footnotesize
\rotate
\tablecaption{GRB used for analysis (GRBs are sorted increasingly by the time of the event) \label{tab:list}}

\tablehead{\colhead{GRB Name} & \multicolumn{1}{p{1.7cm}}{\centering Date (MST)} & \multicolumn{1}{p{1.5cm}}{\centering R.A.\\ Deg., J2000.0} & \multicolumn{1}{p{1.5cm}}{\centering Decl \\ Deg., J2000.0} & \colhead{\parbox{1.0cm}{\centering Redshift}} & \multicolumn{1}{p{1.5cm}}{\centering T start \\ (UTC)} & \multicolumn{1}{p{1.5cm}}{\centering Duration \\ seconds \tablenotemark{a}} & \multicolumn{1}{p{1.5cm}}{\centering Flux $(10^{-5})$ \tablenotemark{b} \\ ph$\cdot$cm$^{-2}\cdot$s$^{-1}$ } & \multicolumn{1}{p{1.5cm}}{\centering Photon \\ index \tablenotemark{b}} & \colhead{\centering TS \tablenotemark{b}} 
}

\startdata
{ 080916C}  & 16 September 2008 & 119.85 & $-$56.60 & 4.35     &
00:12:45.6 & 1775.9 & 7.89$\pm$0.49         & 2.25$\pm$0.06      
& 1140.7\\
{ 090323 } & 23 March 2009 & 190.71 & 17.10  & 3.57     & 00:02:42.6 &
5615.9 & 1.42$\pm$0.19         & 2.28$\pm$0.13              & 205.0  \\
{ 090328}  & 28 March 2009 & 90.67  & $-$42.00 & 0.74     & 09:36:46.5 &
7485.6 & 0.52$\pm$0.07         & 2.10$\pm$0.11              & 253.6  \\
{ 090510}  & 10 May 2009 & 333.55 & $-$26.60 & 0.90     & 00:22:59.9 &
177.8 & 34.11$\pm$2.11         & 2.05$\pm$0.05              & 1234.6 \\
{ 090902B}  & 2 September 2009 & 264.94 & 27.32  & 1.82     & 11:05:08.3
& 749.5 & 14.32$\pm$0.70         & 1.91$\pm$0.04              & 2219.0 \\
{ 090926A}  & 26 September 2009 & 353.40 & $-$66.32 & 2.11     &
04:20:26.9 & 4889.3 & 4.61$\pm$0.28         & 2.08$\pm$0.05     
& 1267.2 \\
{ 091003}  & 3 October 2009 & 251.52 & 36.63  & 0.90     & 04:35:45.5 &
451.6 & 2.17$\pm$0.40         & 2.04$\pm$0.15              & 192.3  \\
{ 100414A} & 14 April 2010 & 192.11 & 8.69   & 1.37     & 02:20:21.9  &
5622.5 & 0.39$\pm$0.07         & 1.77$\pm$0.11              & 188.8  \\
{ 100728A} & 28 July 2010 & 88.76  & $-$15.26 & 1.57     & 02:17:30.6 &
693.7 & 0.65$\pm$0.19         & 1.92$\pm$0.21              & 69.9   \\
{ 110731A} & 31 July 2011 & 280.50 & $-$28.54 & 2.83     & 11:09:29.9 &
561.3 & 3.20$\pm$0.45         & 2.22$\pm$0.12              & 194.3  \\
{ 120624B} & 24 June 2012 & 170.87 & 8.93   & 0.57     & 22:23:53.0 &
1104.3 & 3.86$\pm$0.35         & 2.46$\pm$0.10              & 456.3  \\
{ 120711A} & 11 July 2012 & 94.69  & $-$71.00 & 1.41     & 02:44:53.0 &
5307.2 & 0.56$\pm$0.12         & 1.93$\pm$0.15              & 136.7  \\
{ 130427A} & 27 April 2013 & 173.14 & 27.71  & 0.34     & 07:47:6.0 &
10000 & 4.19$\pm$0.18         & 1.99$\pm$0.03              & 2755.8 \\
{ 130518A} & 18 May 2013 & 355.67 & 47.47  & 2.49     & 13:54:37.0 &
302.9 & 5.38$\pm$0.94         & 2.54$\pm$0.19              & 106.0  \\
{ 130702A} & 2 July 2013 & 217.31 & 15.77  & 0.15     & 00:05:23.0 &
384.5 & 0.04$\pm$0.02         & 1.56$\pm$0.32              & 42.8   \\
{ 130907A} & 7 September 2013 & 215.89 & 45.61  & 1.24     & 21:42:19.0 &
16600 & 0.91$\pm$0.56         & 2.10$\pm$0.46              & 12.8   \\
{ 131108A} & 8 November 2013 & 156.50 & 9.66   & 2.40     & 20:41:55.0 &
1333.5 & 4.52$\pm$0.32         & 2.63$\pm$0.09              & 559.0  \\
{ 131231A} & 31 December 2013 & 10.59  & $-$1.65  & 0.62     & 04:45:16.1
& 5605.6 & 0.34$\pm$0.07         & 1.73$\pm$0.12              & 229.1  \\
{ 141028A} & 28 October 2014 & 322.60 & $-$0.23  & 2.33     & 10:55:03.08
& 414.2 & 2.33$\pm$0.50         & 2.22$\pm$0.21              & 86.7   \\
{ 150314A} & 14 March 2015 & 126.68 & 63.83  & 1.76     & 04:54:50.0 &
250 & 2.24$\pm$0.88         & 2.66$\pm$0.41              & 19.4   \\
{ 150403A} & 3 April 2015 & 311.51 & $-$62.71 & 2.06     & 21:54:10.9 &
1678.3 & 0.22$\pm$0.08         & 1.87$\pm$0.23              & 42.1   \\
{ 150514A} & 14 May 2015 & 74.88  & $-$60.91 & 0.81     & 18:35:05.4 &
600 & 0.09$\pm$0.07         & 1.30$\pm$0.42              & 28.3   \\

\enddata
\label{tab:list}

\tablenotetext{a}{Duration of the GRB considered for our analysis.}
\tablenotetext{b}{Parameters obtained from analysis described in Section~\ref{sec:analysis}}

\end{deluxetable}

\begin{deluxetable}{cccccc}
\tablewidth{0pt}
\footnotesize
\rotate
\tablecaption{Photons detected by the {\it Fermi}-LAT at an optical depth greater than 0.1 \label{tab:nphotons}}

\tablehead{\colhead{GRB Name} & 
\colhead{Redshift} & 
\multicolumn{1}{p{2.4cm}}{\centering Number of photons \\ \cite{finke10} \tablenotemark{a}} &  \multicolumn{1}{p{2.4cm}}{\centering Number of photons \\ \cite{dominguez11} \tablenotemark{b}} &  \multicolumn{1}{p{2.4cm}}{\centering Number of photons \\ \cite{kneiske10} \tablenotemark{c}} &
\multicolumn{1}{p{2.4cm}}{\centering Corresponding energy\\ of photons (GeV) \tablenotemark{d}}}

\startdata
{ 080916C} & 4.35 & 2 & 0 & 2 & 12.4,27.4\\
{ 090323 } & 3.57 & 0 & 0 & 0 & $-$\\
{ 090328}  & 0.74 & 0 & 0 & 0 & $-$\\
{ 090510}  & 0.90 & 0 & 0 & 0 & $-$\\
{ 090902B} & 1.82 & 2 & 0 & 2 & 39.9,21.7\\
{ 090926A} & 2.11 & 1 & 0 & 1 & 19.5\\
{ 091003}  & 0.90 & 0 & 0 & 0 & $-$\\
{ 100414A} & 1.37 & 1 & 0 & 2 & 29.8\\
{ 100728A} & 1.57 & 0 & 0 & 0 & $-$\\
{ 110731A} & 2.83 & 0 & 0 & 0 & $-$\\
{ 120624B} & 0.57 & 0 & 0 & 0 & $-$\\
{ 120711A} & 1.41 & 0 & 0 & 0 & $-$\\
{ 130427A} & 0.34 & 1 & 3 & 2 & 94.1\\
{ 130518A} & 2.49 & 0 & 0 & 0 & $-$\\
{ 130702A} & 0.15 & 0 & 0 & 0 & $-$\\
{ 130907A} & 1.24 & 1 & 0 & 1 & 50.9\\
{ 131108A} & 2.40 & 0 & 0 & 0 & $-$\\
{ 131231A} & 0.62 & 0 & 0 & 1 & $-$\\
{ 141028A} & 2.33 & 0 & 0 & 0 & $-$\\
{ 150314A} & 1.76 & 0 & 0 & 0 & $-$\\
{ 150403A} & 2.06 & 0 & 0 & 0 & $-$\\
{ 150514A} & 0.81 & 0 & 0 & 0 & $-$\\

\enddata

\tablenotetext{a}{Number of LAT Photons detected at $\tau>0.1$ (obtained using EBL model \cite{finke10}$-model C$).}
\tablenotetext{b}{Upper limit of the number of LAT Photons detected at $\tau>0.1$ (obtained using EBL model \cite{dominguez11}).}
\tablenotetext{c}{{}Lower limit of the number of LAT Photons detected at $\tau>0.1$ (obtained using EBL model \cite{kneiske10}).}
\tablenotetext{d}{Energy of the photons detected at $\tau>0.1$ (obtained using EBL model \cite{finke10}$-model C$). }

\end{deluxetable}

\begin{deluxetable}{l|ccc|cc|c}
\tablewidth{0pt}
\tablecaption{Joint-likelihood results for different EBL models using GRB sources.}
\label{tab:grbresults}
\tablehead{\colhead{Model} &
TS$_{0}$\tablenotemark{a} &
p$_{0}$\tablenotemark{b} & 
$b$\tablenotemark{c}  & 
TS$_{1}$\tablenotemark{d} &
p$_{1}$\tablenotemark{b} &
$\Delta$TS\tablenotemark{e}
}

\startdata

{\it \cite{kneiske04} -- high UV} & 6.5 & 2.55 & $0.43_{-0.28}^{+0.24}$  & 3.5 & 1.87 & 3.0\\ 
{\it \cite{kneiske04} -- best-fit} & 7.4 & 2.72 & $0.80_{-0.61}^{+0.51}$  & 0.1 & 0.32 & 7.3\\ 
{\it \cite{primack05}} & 4.7 & 2.17 & $0.51_{-0.38}^{+0.34}$  & 1.5 & 1.22 & 3.2\\ 
{\it \cite{gilmore09}} & 7.1 & 2.66 & $1.25_{-0.95}^{+0.82}$  & 0.1 & 0.32 & 7.0\\
{\it \cite{finke10} --  model C}& 7.7 & 2.77 & $1.27_{-0.99}^{+0.84}$  & 0.1 & 0.32 & 7.6\\ 
{\it \cite{kneiske10}} & 7.4 & 2.72 & $1.29_{-0.95}^{+0.80}$  & 0.2 & 0.45 & 7.2\\ 
{\it \cite{dominguez11}} & 8.0 &  2.83 & $2.21_{-1.83}^{+1.48}$  & 1.0 & 1.00 & 7.0\\ 
{\it \cite{gilmore12} -- fixed}& 7.3 & 2.70 & $1.43_{-1.13}^{+0.93}$  & 0.3 & 0.55 & 7.0\\ 
{\it \cite{gilmore12} -- fiducial}& 6.5 & 2.55 & $0.63_{-0.46}^{+0.40}$  & 0.7 & 0.84 & 5.8\\
{\it \cite{helgason12}}& 7.2 & 2.68 & $1.44_{-1.18}^{+0.95}$  & 0.3 & 0.55 & 6.9\\
{\it \cite{scully14} -- Low Opacity}& 6.9 & 2.62 & $1.16_{-0.79}^{+0.69}$  & 0.1 & 0.32 & 6.8\\ 
{\it \cite{scully14} -- High Opacity}& 6.7 & 2.59 & $0.42_{-0.29}^{+0.25}$  & 3.3 & 1.82 & 3.4\\ 
{\it \cite{inoue13} }& 6.4 & 2.53 & $0.72_{-0.50}^{+0.43}$  & 0.4 & 0.63 & 6.0\\

\enddata
\label{tab:grbresults}
\tablenotetext{a}{TS obtained from the comparison of 
the null hypothesis ($b$=0) with the likelihood obtained with best-fit value for $b$.}
\tablenotetext{b}{The $p_0$ and $p_1$ values are denoted in units of standard deviation of a normal Gaussian distribution.}
\tablenotetext{c}{This column lists the best-fit values 
and 1\,$\sigma$ confidence ranges for the opacity scaling factor.}
\tablenotetext{d}{Here the compatibility 
of the predictions of EBL models with the {\it Fermi}
observations is shown ($b$=1 case constitutes the null hypothesis). Large values mean less likely to be compatible.}
\tablenotetext{e}{$\Delta$TS= TS$_{0}-$ TS$_{1}$}

\end{deluxetable}

\begin{deluxetable}{l|ccc|cc|c}
\tablewidth{0pt}
\tablecaption{Combined results of GRB and BL Lac sources for different EBL models.}
\tablehead{\colhead{Model\tablenotemark{a}} &
TS$_{0}$ \tablenotemark{b} &
p$_{0}$\tablenotemark{c} &
$b$\tablenotemark{d}          & 
TS$_{1}$ \tablenotemark{e} &
p$_{1}$\tablenotemark{c} &
$\Delta$TS\tablenotemark{f}
}
\startdata
{\it \cite{kneiske04} -- high UV} & 32.5 & 5.70 & 0.38$\pm$0.08  & 38.3 & 6.19 & -5.8\\ 
{\it \cite{kneiske04} -- best-fit} & 41.0 & 6.40 & 0.54$\pm$0.12  & 10.4 & 3.22 & 30.6\\ 
{\it \cite{primack05}}& 35.0 & 5.91 & 0.73$\pm$0.14  & 5.0 & 2.23 & 30.0\\ 
{\it \cite{gilmore09}} & 40.7 & 6.38 & 0.99$\pm$0.21  & 0.1 & 0.32 & 40.6\\ 
{\it \cite{finke10} --  model C}& 41.3 & 6.43 & 0.88$\pm$0.22 & 0.3 & 0.55 & 41.0\\ 
{\it \cite{kneiske10}} & 39.9 & 6.32 & 0.92$\pm$0.18  & 0.19 & 0.44 & 39.7\\ 
{\it \cite{dominguez11}} & 42.8 & 6.54 & 1.04$\pm$0.23  & 0.04 & 0.20 & 42.8\\ 
{\it \cite{gilmore12} -- fixed} & 40.7 & 6.38 & 1.04$\pm$0.22  & 0.04 & 0.20 & 40.7\\ 
{\it \cite{gilmore12} -- fiducial} & 40.1 & 6.33 & 0.92$\pm$0.20  & 0.16 & 0.40 & 39.9\\
\enddata
\label{tab:cresults}
\tablenotetext{a}{Only models common to \cite{ebl12} and our analysis are listed here.}
\tablenotetext{b}{Same as table 2 but combined TS obtained from GRB and BL Lac observations}
\tablenotetext{c}{The $p_0$ and $p_1$ values are denoted in units of standard deviation of a normal Gaussian distribution.}
\tablenotetext{d}{Maximum likelihood values and uncertainty obtained by performing a weighted average of GRB and BL Lac data}
\tablenotetext{e}{Same as table 2 but combined TS obtained from GRB and BL Lac observation.}
\tablenotetext{f}{$\Delta$TS= TS$_{0}-$ TS$_{1}$}

\end{deluxetable}

\end{document}